\documentclass[aps,prd,preprint,superscriptaddress,showpacs]{revtex4}

\usepackage{amsfonts}
\usepackage{mathrsfs}
\usepackage{amssymb}
\usepackage{amsmath}
\usepackage{graphicx,epsfig}

\def\bwt{\begin{widetext}}

\def\ewt{\end{widetext}}

\def\be{\begin{equation}}

\def\ee{\end{equation}}

\def\bea{\begin{eqnarray}}

\def\eea{\end{eqnarray}}

\def\bean{\begin{eqnarray*}}

\def\eean{\end{eqnarray*}}

\def\bary{\begin{array}}

\def\eary{\end{array}}

\def\bit{\begin{itemize}}

\def\eit{\end{itemize}}

\def\su5u1{SU(5) \times U(1)}

\def\fsu5u1{SU(5) \times U(1)'}

\def\so10{SO(10)}

\def\sq20{SO(10) \times SO(10)}

\usepackage{amssymb}

\begin{document}

\setlength{\parskip}{0cm}

\title{ Canonical Gauge Coupling Unification in the Standard Model with
High-Scale Supersymmetry Breaking}

\author{Yun-Jie Huo}

\affiliation{Key Laboratory of Frontiers in Theoretical Physics,
      Institute of Theoretical Physics, Chinese Academy of Sciences,
Beijing 100190, P. R. China }

\author{Tianjun Li}

\affiliation{Key Laboratory of Frontiers in Theoretical Physics,
      Institute of Theoretical Physics, Chinese Academy of Sciences,
Beijing 100190, P. R. China }

\affiliation{George P. and Cynthia W. Mitchell Institute for
Fundamental Physics, Texas A$\&$M University, College Station, TX
77843, USA }

\author{Dimitri V. Nanopoulos}

\affiliation{George P. and Cynthia W. Mitchell Institute for
Fundamental Physics,
 Texas A$\&$M University, College Station, TX 77843, USA }

\affiliation{Astroparticle Physics Group,
Houston Advanced Research Center (HARC),
Mitchell Campus, Woodlands, TX 77381, USA}

\affiliation{Academy of Athens, Division of Natural Sciences,
 28 Panepistimiou Avenue, Athens 10679, Greece }



\begin{abstract}

Inspired by the string landscape and 
the unified gauge coupling relation in 
 the F-theory Grand Unified Theories (GUTs) and GUTs with
suitable high-dimensional operators, we study the 
canonical gauge coupling unification and Higgs boson mass
in the Standard Model (SM) with high-scale supersymmetry breaking. 
In the SM with GUT-scale supersymmetry breaking, we achieve the
 gauge coupling unification 
 at about $5.3\times 10^{13}~{\rm GeV}$, and the
Higgs boson mass is predicted to range from 130 GeV to 147 GeV.
In the SM with supersymmetry breaking scale from $10^4~{\rm GeV}$
to $5.3\times 10^{13}~{\rm GeV}$, gauge coupling unification
can always be realized and the corresponding
GUT scale $M_U$ is  from $10^{16}~{\rm GeV}$ to 
$5.3\times 10^{13}~{\rm GeV}$, respectively. Also,
we obtain the Higgs
boson mass from 114.4 GeV to $147$ GeV.
Moreover, the discrepancies among the SM gauge couplings
at the GUT scale are less than about 4-6\%.
Furthermore, we present the $SU(5)$ and $SO(10)$ models from the 
F-theory model building and orbifold constructions, and show
that we do not have the dimension-five and dimension-six
proton decay problems even if $M_U \le 5\times 10^{15}~{\rm GeV}$.

\end{abstract}


\preprint{ACT-13-10, MIFPA-10-49}

\maketitle



\section{Introduction}

It is well-known that there might exist an enormous ``landscape'' for long-lived
metastable string/M theory vacua where the
moduli can be stabilized and supersymmetry may be broken
in the string models with flux compactifications~\cite{String}. 
Applying the ``weak anthropic principle''~\cite{Weinberg}, the 
string landscape proposal might
provide the first concrete solution to the cosmological constant problem,
 and  it may  address the gauge hierarchy problem in the Standard Model (SM).  
Notably, the supersymmetry breaking scale 
can be high if there exist many supersymmetry breaking parameters or many hidden
sectors~\cite{HSUSY,NASD}.  Although there is no definite conclusion whether
the string landscape predicts high-scale or TeV-scale supersymmetry 
breaking~\cite{HSUSY},
it is interesting to study the models with high-scale supersymmetry
breaking due to the turn on of the 
Large Hadron Collider 
(LHC)~\cite{NASD, Barger:2004sf, Barger:2005gn, Barger:2005qy, Hall:2009nd}.

Assuming that supersymmetry is indeed broken at a high scale, we can classify
the supersymmetry breaking scale as follows~\cite{Barger:2004sf}:
(1)~the string scale or grand unification scale;
(2)~an intermediate scale; and (3)~the TeV scale. 
We do not consider the TeV-scale supersymmetry here since it  
has been studied extensively during the  last thirty years. However,
we would like to emphasize that for high-scale supersymmetry breaking,
most of the problems associated with some 
low energy supersymmetric models, for example,  
excessive flavor and CP violations, dimension-five fast proton decay and the stringent 
constraints on the lightest CP-even neutral 
Higgs boson mass, may be solved automatically. 

If supersymmetry is broken at the high scale, the  minimal model at the low energy
is the Standard model. The SM explains existing experimental data very well,
 including electroweak precision tests. Moreover,
we can easily incorporate aspects of physics beyond the SM through small 
variations, for example,  dark matter, dark energy,
atmospheric and solar neutrino oscillations,
 baryon asymmetry, and inflation~\cite{Davoudiasl:2004be}. 
Also, the SM fermion masses and mixings can be
explained via the Froggatt-Nielsen mechanism~\cite{FN}.
However, there are still some limitations of the SM,
for example, the lack of explanation of gauge coupling unification and charge
quantization~\cite{Barger:2005gn, Barger:2005qy}. 

Charge quantization can easily be realized by
embedding the SM into the Grand Unified Theories (GUTs). 
Anticipating that the Higgs particle might be the only
new physics observed at the LHC, thus confirming the SM 
as the low energy effective theory, we should
reconsider gauge coupling unification in the SM.
Previously, the generic gauge coupling unification 
can be defined by
\begin{eqnarray}
k_Y g_Y^2 =  g_2^2 =  g_3^2  ~,
\end{eqnarray}
where $k_Y$ is the normalization constant for 
the $U(1)_Y$ hypercharge interaction, and
$g_Y$, $g_2$, and $g_3$ are the gauge couplings for
the $U(1)_Y$, $SU(2)_L$, and $SU(3)_C$ gauge groups, respectively.
However, it is well-known that
gauge coupling unification cannot be achieved in the SM with
 canonical $U(1)_Y$ normalization, 
{\it i.e.}, the Georgi-Glashow $SU(5)$ 
normalization with $k_Y=5/3$~\cite{Ellis:1990zq}.
Interestingly, it was shown that gauge coupling unification
can be realized in the non-canonical $U(1)_Y$ normalization
 with $k_Y=4/3$~\cite{Barger:2005gn, Barger:2005qy}. 
The orbifold GUTs with such
$U(1)_Y$ normalization have been constructed as well.
The key question remains: can we realize the gauge coupling
unification in the SM with canonical $U(1)_Y$ normalization?

During the last a few years, GUTs have been constructed 
locally in the F-theory model building~\cite{Vafa:1996xn,
Donagi:2008ca, Beasley:2008dc, Beasley:2008kw, Donagi:2008kj,
Font:2008id, Jiang:2009zza, Blumenhagen:2008aw, Jiang:2009za,
Li:2009cy}.
A brand new feature is that the $SU(5)$ gauge symmetry
can be broken down to the SM gauge symmetry
by turning on $U(1)_Y$ 
flux~\cite{Beasley:2008dc, Beasley:2008kw, Li:2009cy}, 
and the $SO(10)$  gauge 
symmetry can be broken down to the $SU(5)\times U(1)_X$
and $SU(3)_C\times SU(2)_L\times SU(2)_R\times U(1)_{B-L}$
gauge symmetries by turning on the $U(1)_X$ and $U(1)_{B-L}$
fluxes, respectively~\cite{Beasley:2008dc, Beasley:2008kw, 
Jiang:2009zza, Jiang:2009za, Font:2008id, Li:2009cy}. 
It has been shown that the gauge kinetic functions receive
the corrections from $U(1)$ fluxes~\cite{Donagi:2008kj, 
Blumenhagen:2008aw, Jiang:2009za, Li:2009cy}. In particular,
in the $SU(5)$ models with $U(1)_Y$ flux~\cite{Donagi:2008kj, 
Blumenhagen:2008aw} and in the $SO(10)$ models 
with $U(1)_{B-L}$ flux~\cite{Li:2009cy}, the SM gauge couplings
 at the GUT scale satisfy the following condition
\begin{eqnarray}
{{1}\over {\alpha_2}} - {{1}\over {\alpha_3}} 
~=~{5\over 3} \left( {{1}\over {\alpha_1}} 
- {{1}\over {\alpha_3}} \right) ~,~\,
\label{GCRelation}
\end{eqnarray}
where $\alpha_1=5 \alpha_Y/3$, $\alpha_Y=g_Y^2/4\pi$,
and $\alpha_j = g_j^2/4\pi$ for $j=2,~3$. 
In other words, the gauge coupling unification scale
$M_U$ is defined by Eq.~(\ref{GCRelation}). Especially,
we have canonical $U(1)_Y$ normalization here. Moreover,
the above gauge coupling relation at the GUT scale can
be realized in the four-dimensional GUTs with suitable high-dimensional 
operators~\cite{Hill:1983xh, Shafi:1983gz, Ellis:1985jn, Li:2010xr}
 and in the orbifold GUTs~\cite{kawa, GAFF, LHYN, AHJMR, Li:2001qs, 
Dermisek:2001hp, Li:2001tx} with similar high-dimensional
operators on the 3-branes at the fixed points where the
complete GUT gauge symmetries are 
preserved. We emphasize that the above gauge coupling relation
at the GUT scale was first given in Ref.~\cite{Ellis:1985jn}.

In this paper, considering high-scale supersymmetry breaking
inspired by the string landscape, 
we shall study the gauge coupling unification
in the SM where the GUT-scale gauge coupling relation 
is given by Eq.~(\ref{GCRelation}). In the SM with
GUT-scale supersymmetry breaking, the SM gauge couplings
are unified at about $5.3\times 10^{13}~{\rm GeV}$.
In the SM with supersymmetry breaking scale from $10^4~{\rm GeV}$
to $5.3\times 10^{13}~{\rm GeV}$, gauge coupling unification
can always be realized, and we obtain the corresponding
GUT scale $M_U$ from $10^{16}~{\rm GeV}$ to 
$5.3\times 10^{13}~{\rm GeV}$, respectively.
Also, the discrepancies among the SM gauge couplings
at the GUT scale are less than about 4-6\%.
Moreover, we calculate the SM Higgs boson mass. 
In the SM with GUT-scale supersymmetry breaking, the
Higgs boson mass is predicted to range from 130 GeV to 147 GeV.
And in the SM with supersymmetry breaking scale from $10^4~{\rm GeV}$
to $5.3\times 10^{13}~{\rm GeV}$, we obtain the Higgs
boson mass from 114.4 GeV to $147$ GeV
where the low bound on the SM Higgs boson
mass from the LEP experiment~\cite{Barate:2003sz} has been included.
Furthermore, we present the $SU(5)$ and $SO(10)$ models from the 
F-theory model building and orbifold constructions, and show
that there are no dimension-five and dimension-six
proton decay problems even if $M_U \le 5\times 10^{15}~{\rm GeV}$.

This paper is organized as follows. In Section II, we study
the gauge coupling unification in the SM with high-scale 
supersymmetry breaking. In Section III, we 
consider the Higgs boson masses. We present the concrete
$SU(5)$ and $SO(10)$ models without proton decay 
problems in Section IV. And our conclusion
is given in Section V.



\section{Gauge Coupling Unification}



For simplicity, we consider the universal high-scale supersymmetry breaking. 
Above the universal supersymmetry breaking scale $M_S$, we consider 
the supersymmetric SM. Following the procedures in Ref.~\cite{Barger:2005qy} 
where all 
the relevant renormalization group equations
(RGEs) are given, we consider the two-loop RGE running for 
the SM gauge couplings, and one-loop RGE running for the
SM fermion Yukawa couplings.

In numerical calculations, we choose the top quark pole mass 
$M_t = 173.1\pm 1.3 $ GeV~\cite{:2009ec}, 
and the strong coupling constant
$\alpha_3(M_Z) = 0.1184 \pm 0.0007$~\cite{Nakamura:2010zzi}, 
where $M_Z$ is the $Z$ boson mass. Also, the
fine structure constant $\alpha_{EM}$, weak mixing angle $\theta_W$ and 
Higgs vacuum expectation value (VEV) $v$ at $M_Z$ are taken as 
follows~\cite{Nakamura:2010zzi}
\bea
\alpha^{-1}_{EM}(M_Z) = 128.91 \,, ~~~
\sin^2\theta_W(M_Z) = 0.23116\,, ~~~
v = 174.10\,{\rm GeV}\,.
\eea

\begin{figure}[htb]
\centering
\includegraphics[width=13cm]{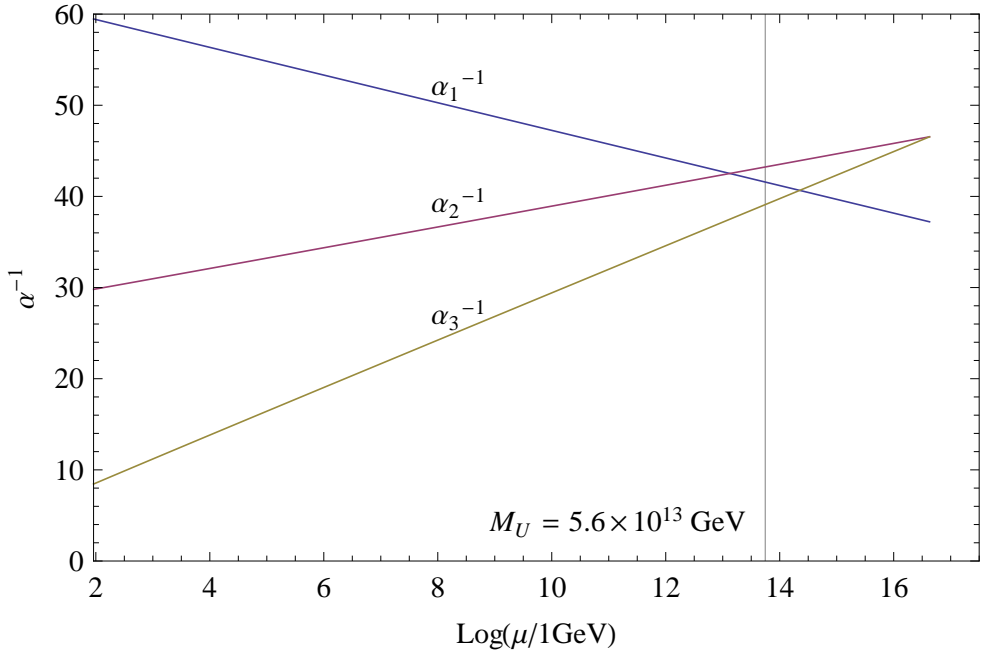}
\caption{Canonical gauge coupling unification in the SM where
the gauge coupling unification scale
$M_U$ is defined by Eq.~(\ref{GCRelation}).
}
\label{gcusm}
\end{figure}

First, we consider the GUT-scale universal supersymmetry breaking,
{\it i.e.}, we only have the SM below the GUT scale. 
With the GUT-scale gauge coupling relation in Eq.~({\ref{GCRelation}),
we present the gauge coupling unification in Fig.~\ref{gcusm},
and find that the unification scale is about 
$5.3\times 10^{13}~{\rm GeV}$. Next, we consider 
the intermediate-scale universal supersymmetry breaking. 
Interestingly, gauge coupling unification can always be
realized. In Fig.~\ref{muvsms},
we present the GUT scale for the universal supersymmetry breaking
scale $M_S$ from $10^4~{\rm GeV}$ to $5.3\times10^{13}~{\rm GeV}$. 
The GUT scale decreases when the supersymmetry breaking
scale increases. Moreover, the GUT scale
varies from $10^{16}~{\rm GeV}$ to
$5.3\times10^{13}~{\rm GeV}$ for the supersymmetry breaking
scale from  $10^4~{\rm GeV}$
to $5.3\times10^{13}~{\rm GeV}$, respectively.
Moreover, the GUT scale is almost independent on the
mixing parameter $\tan\beta$, which is defined in the 
first paragraph in the next Section.

\begin{figure}[htb]
\centering
\includegraphics[width=13cm]{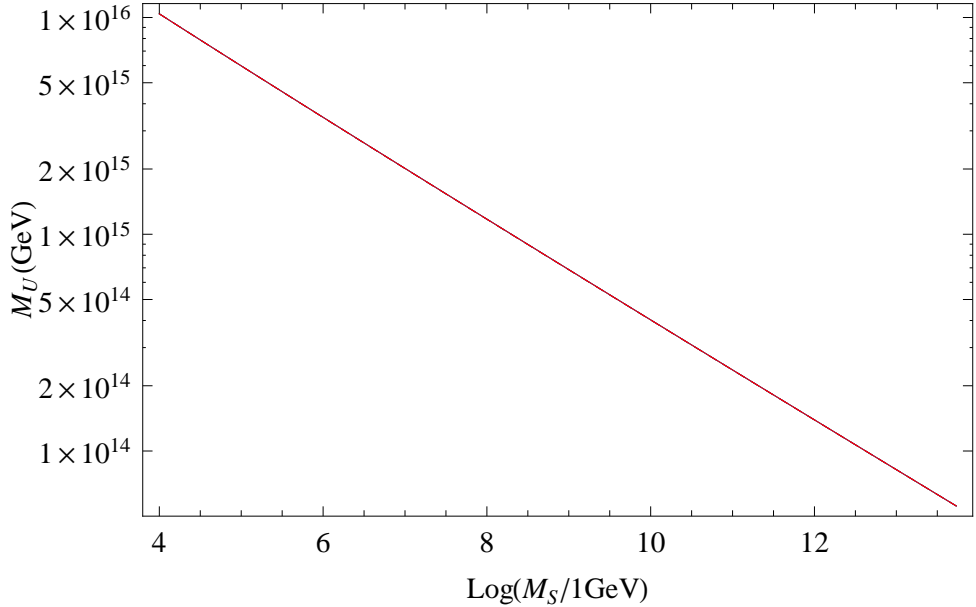}
\caption{The GUT scale $M_U$ versus the universal supersymmetry breaking
scale $M_S$. We consider $\tan\beta$ =
3 (dotted line) and 35 (solid line), and $M_t$ = 171.8~GeV, 173.1~GeV,
 174.4~GeV. The results  for different
cases are roughly the same. }
\label{muvsms}
\end{figure}


To demonstrate that the deviations from the complete gauge coupling
universality are still modest, we 
 study the discrepancies among the SM gauge couplings at
the GUT scale by defining two parameters $\delta_+$ and $\delta_-$ 
at the GUT scale
\begin{equation}
\delta_+=\frac{\alpha_2^{-1}-\alpha_1^{-1}}{\alpha_1^{-1}}~,~~~~~~~
~~~~~\delta_-=\frac{\alpha_3^{-1}-\alpha_1^{-1}}{\alpha_1^{-1}}~.~\,
\end{equation}
In Fig.~\ref{dvsms}, we present $\delta_+$ and $\delta_-$  
for the supersymmetry breaking
scale from $10^4~{\rm GeV}$ to $5.3\times10^{13}~{\rm GeV}$. 
We find that $\delta_+$ and $|\delta_-|$ increase when  
the supersymmetry breaking scale
$M_S$ increases. Also,   $\delta_+$ and $|\delta_-|$ 
are smaller than 4\% and 6\%, respectively.
Similar to the GUT scale, $\delta_+$ and $\delta_-$ 
are almost independent on $\tan\beta$ as well.
 Thus, these discrepancies among the SM gauge
couplings at the GUT scale are indeed  small.

\begin{figure}[htb]
\centering
\includegraphics[width=13cm]{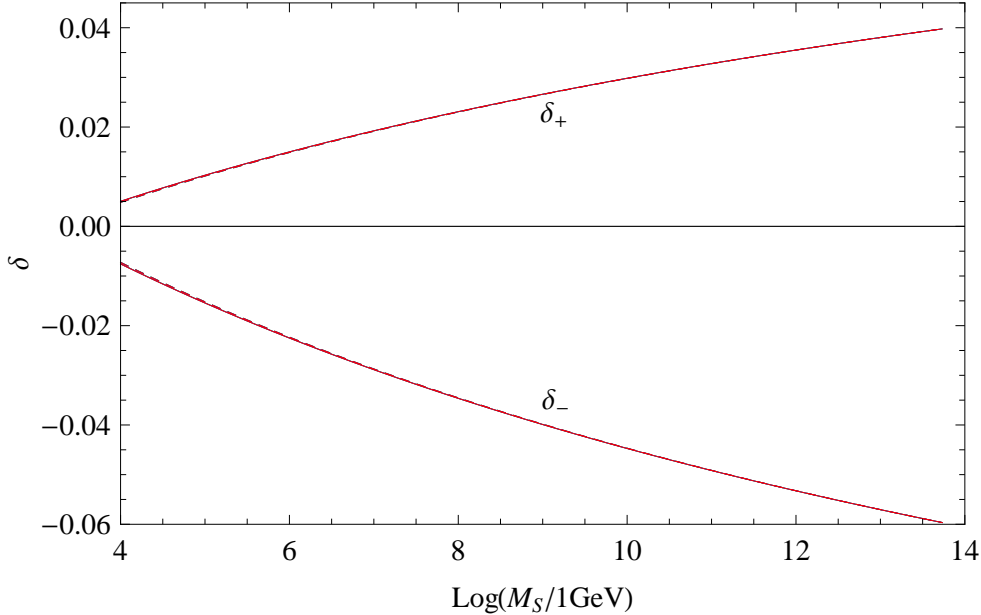}
\caption{ $\delta_+$ and $\delta_-$ versus the universal supersymmetry
breaking scale $M_S$. We consider
$\tan\beta$ = 3 (dotted line) and 35 (solid line), and $M_t$ = 171.8~GeV,
 173.1~GeV, and 174.4~GeV. The results for different
cases are roughly the same.}
\label{dvsms}
\end{figure}




\section{Higgs Boson Mass}



If the Higgs particle is the only new physics discovered at the LHC
and then the  SM is  confirmed as the low energy effective theory,
the Higgs boson mass is one of the most important parameters.
Above the supersymmetry breaking scale, we have supersymmetric SMs.
There generically exists one pair of Higgs doublets $H_u$ and $H_d$,
which give masses to the up-type quarks and down-type quarks/charged
leptons, respectively. Below the supersymmetry breaking scale,
we only have the SM. Let us define the SM Higgs doublet
 $H$ as $H\equiv-\cos\beta
i\sigma_2H_d^\ast+\sin\beta H_u$, where $\sigma_2$ is the second
Pauli matrix and $\tan\beta$ is a mixing 
parameter~\cite{NASD, Barger:2004sf, Barger:2005gn}. For simplicity,
we assume the gauginos, squarks, Higgsinos, and the other
combination of the scalar Higgs doublets $\sin\beta
i\sigma_2H_d^\ast+\cos\beta H_u$ have the universal supersymmetry
breaking soft mass $M_S$. We first assume that supersymmetry is
broken at the GUT scale $M_U$, {\it i.e.},  $M_S \simeq M_U$. 
And then we assume that supersymmetry is broken at 
the intermediate scale,  {\it i.e.},
 below the GUT scale but higher than the electroweak
scale, such as between $10^4$ GeV and $M_U$. 

We consider the supersymmetry breaking scale $M_S$ from $10^4~{\rm GeV}$ 
to the SM unification scale $5.3\times 10^{13}~{\rm GeV}$.
At the supersymmetry breaking scale, we can calculate the Higgs boson 
quartic coupling
$\lambda$~\cite{NASD, Barger:2004sf, Barger:2005gn}
\begin{equation}
\lambda(M_S)=\frac{g_1^2(M_S)+k_Yg_2^2(M_S)}{4k_Y}\cos^22\beta,
\end{equation}
where $k_Y=5/3$, and then evolve it down to the Higgs boson mass scale.
The one-loop RGE for the quartic coupling is given in Ref.~\cite{Barger:2005qy} as well.
To predict the SM Higgs boson mass, we consider the two-loop RGE running for 
the SM gauge couplings, and one-loop RGE running for the
SM fermion Yukawa couplings and Higgs quartic coupling.
Using the one-loop effective Higgs potential with top quark
radiative corrections, we calculate the Higgs boson mass by
minimizing the effective potential
\begin{equation}
V_{eff}=m_h^2H^\dagger H+\frac{\lambda}{2!}(H^\dagger
H)^2-\frac{3}{16\pi^2}h_t^4(H^\dagger
H)^2\left[\log\frac{h_t^2(H^\dagger H)}{Q^2}-\frac{3}{2}\right],
\end{equation}
where $m_h^2$ is the squared Higgs boson mass, 
$h_t$ is the top quark Yukawa coupling from $m_t=h_tv$,
 and the scale $Q$ is chosen to be at the Higgs boson
mass. For the $\overline{MS}$ top quark mass $m_t$, we use the
two-loop corrected value, which is related to the top quark pole
mass $M_t$ by~\cite{Melnikov:2000qh}
\begin{eqnarray}
M_t&=&m_t(m_t)\left\{1+\frac{4\alpha_3(m_t)}{3\pi}+\left[13.4434-1.0414\sum_{k=1}^5(1-\frac{4}{3}\frac{m_k}{m_t})\right]\left[\frac{\alpha_3(m_t)}{\pi}\right]^2 \right\}~,~
\end{eqnarray}
where $m_k$ denotes the other quark mass. Also, the two-loop 
RGE running for $\alpha_3$ has
been used.

\begin{figure}[htb]
\centering
\includegraphics[width=13cm]{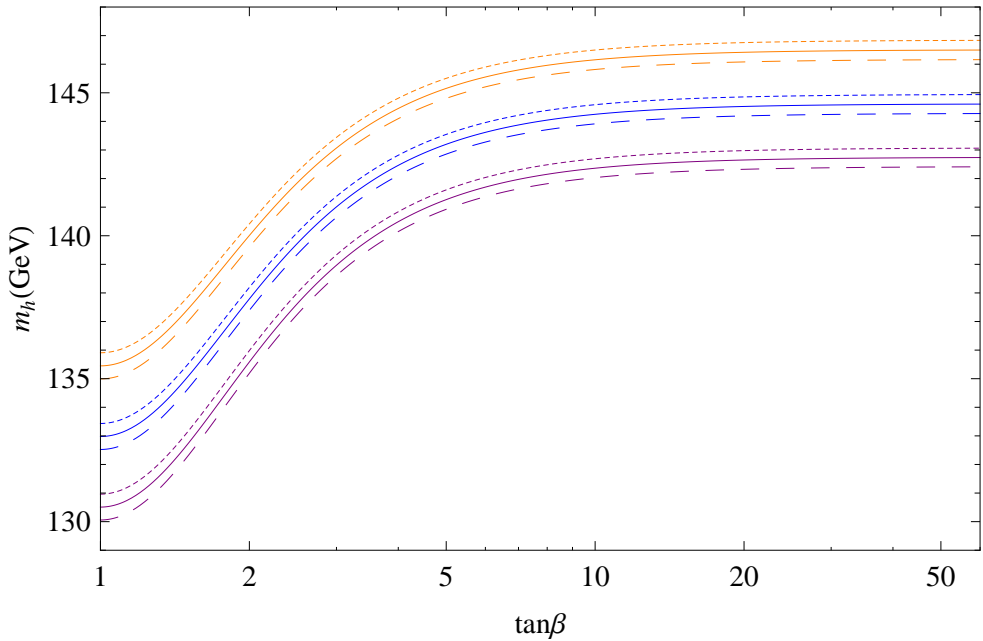}
\caption{ The predicted Higgs boson mass versus $\tan\beta$ in the SM with
GUT scale supersymmetry breaking. The top (orange) three curves are
for $M_t+\delta M_t$, the bottom (purple) $M_t-\delta M_t$, and the
middle (blue) $M_t$. The dotted curves are for
$\alpha_3-\delta\alpha_3$, the dash ones for
$\alpha_3+\delta\alpha_3$, and the solid ones for
$\alpha_3$. Here, we choose $M_t=173.1~{\rm GeV}$ and $\delta M_t=1.3~{\rm GeV}$.}
\label{mhvstb}
\end{figure}


For the SM with GUT-scale supersymmetry breaking, the predicted
Higgs boson mass is shown as a function of $\tan\beta$ for different
$M_t$ and $\alpha_3$ in Fig.~\ref{mhvstb}. 
When we increase top quark mass or decrease strong coupling,
the predicted Higgs boson mass will increase.
If we vary $M_t$ and
$\alpha_s$ within their $1\sigma$ range, and $\tan\beta$ from 1 to
60, the predicted Higgs boson mass will range from $130~{\rm GeV}$ to 
$147~{\rm GeV}$. Moreover, focussing 
on the high-scale supersymmetry breaking around $10^{14}~{\rm GeV}$,
Hall and Nomura 
made a very fine prediction for the Higgs boson mass from $128~{\rm GeV}$ to 
$141~{\rm GeV}$~\cite{Hall:2009nd}. Thus, our predicted Higgs boson masses are
a little bit larger than their results. Concretely speaking,
the discrepancy between our low bound
and their low bound is about 1.5\% while the discrepancy between our upper bound
and their upper bound is about 4\%. Because the inputs for the top quark
mass are the same, it seems to us that these discrepancies may be due to the
following three reasons: (1) Our supersymmetry
breaking scale is $5.3\times10^{13}~{\rm GeV}$ while their 
 supersymmetry breaking scale is $4\times10^{14}~{\rm GeV}$, thus, 
the boundary conditions are different. (2) For the SM fermion Yukawa couplings and 
Higgs quartic coupling, we consider the
one-loop RGE running while they considered the two-loop RGE
running. (3) We consider $\tan\beta$ from 1 to 60 while they
considered  $\tan\beta$ from 1 to 10. Although each of these effects is
small, we may understand the discrepancies by summing up all
these effects.

In Fig.~\ref{mhvsms}, we present the Higgs boson mass for the 
intermediate-scale  supersymmetry breaking. Generically, 
the predicted Higgs boson mass will increase when supersymmetry
breaking scale increases.
For supersymmetry breaking scale $M_S$
varying from $10^4~{\rm GeV}$ to $5.3\times10^{13}~{\rm GeV}$, and $\tan\beta$
between 3 and 35, $M_t$ within its $1\sigma$ range, the predicted
Higgs boson mass will range from $114.4~{\rm GeV}$ to $146~{\rm GeV}$,
where the low bound on the SM Higgs boson mass
from the LEP experiment~\cite{Barate:2003sz} has been included.
If we also vary $\alpha_3$ within its $1\sigma$ range, the predicted
Higgs boson mass will range from 
$114.4~{\rm GeV}$ to $147~{\rm GeV}$.

\begin{figure}[htb]
\centering
\includegraphics[width=13cm]{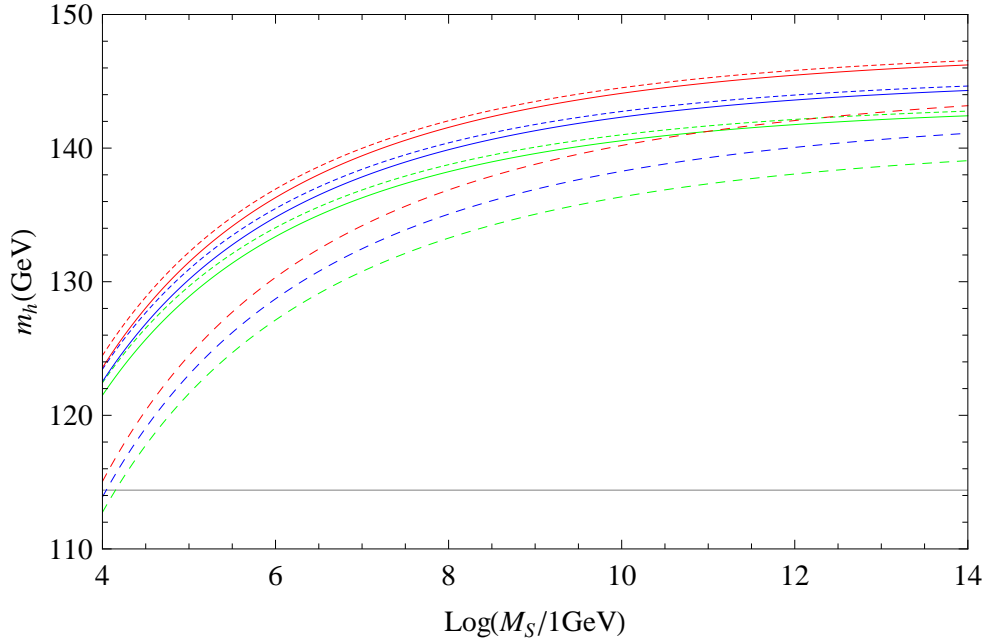}
\caption{The predicted Higgs boson mass versus $M_S$ in the SM with
high-scale supersymmetry breaking. The top (red) two curves are
for $M_t+\delta M_t$, the bottom (green) $M_t-\delta M_t$, and the
middle (blue) $M_t$. 
The dash curves are for $\tan\beta=3$, the
solid ones for $\tan\beta=10$, and the dotted ones for  $\tan\beta=35$.
The horizontal line is the LEP low bound 114.4 GeV.}
\label{mhvsms}
\end{figure}





\section{F-Theory GUTs and Orbifold GUTs}

Because the GUT scale in our models can be as small as 
$5.3\times10^{13}~{\rm GeV}$,
we might have dimension-five and dimension-six proton decay problems.
In this paper, we shall consider the $SU(5)$ and $SO(10)$ models
from the local F-theory constructions and
the orbifold constructions, where these proton decay problems can be solved.
In particular, the GUT-scale gauge coupling relation 
 given by Eq.~(\ref{GCRelation}) can be realized.

Let us explain our convention. 
In the supersymmetric SMs, we denote the left-handed quark doublets,
right-handed up-type quarks, right-handed down-type quarks,
 left-handed lepton doublets, right-handed neutrinos, and 
right-handed charged leptons as $Q_i$, $U_i^c$, $D_i^c$, $L_i$, $N_i^c$,
and $E_i^c$, respectively.
In the $SU(5)$ models, the SM fermions form $\mathbf{10}_i=(Q_i,~U_i^c,~E_i^c)$
and $\mathbf{\overline{5}}_i=(D_i^c,~L_i)$ and $\mathbf{1}_i=N_i^c$ representations.
The Higgs fields form $\mathbf{{5}}_H=(T_u,~H_u)$ and 
$\mathbf{\overline{5}}_{\overline{H}}=(T_d,~H_d)$ representations, where
$T_u$ and $T_d$ are the colored Higgs fields.
In the $SO(10)$ models, one family of the SM fermions form a spinor $\mathbf{16}_i$
representation, and all the Higgs fields form a 
$\mathbf{10}_H=(T_u,~H_u,~T_d,~H_d)$
representation.

First, we briefly review the proton decay. The  
dimension-five proton decays arise from the color-Higgsino exchanges.
In  $SU(5)$ and $SO(10)$ models, we have the following superpotential in 
terms of the SM fermions
\begin{eqnarray}
W &=& y_u^{ij} (Q_i Q_j +  2 U^c_i E^c_j ) T_u + y_{de}^{ij} 
(2 Q_i L_j + 2 U^c_i D^c_j)T_d + M_T T_u T_d ~,~\,
\label{PD-D5}
\end{eqnarray}
where $y_u^{ij} $ are the Yukawa couplings for the
up-type quarks, and  $y_{de}^{ij}$ are the Yukawa couplings for 
the down-type quarks and charged leptons.
In $SO(10)$ models, we shall have $y_u^{ij}= y_{de}^{ij}$ as well.
The dimension-five proton decay operators are obtained after
we integrate out the heavy colored Higgs fields $T_u$ and
$T_d$. The corresponding proton partial lifetime from dimension-five proton decay
is proportional to $M_T^2 M_S^2$,
and we require  $M_T M_S ~\ge~ 10^{20}~{\rm GeV}^2$ from the
current experimental bounds~\cite{Hisano:1992jj, Murayama:2001ur}.

The dimension-six proton decay operators are obtained after we integrate
out the heavy gauge boson fields. In $SU(5)$ models, we have two kinds
of operators $\mathbf{10^*}_i \mathbf{10}_i \mathbf{10^*}_j \mathbf{10}_j$ and 
$\mathbf{10^*}_i \mathbf{10}_i \mathbf{5^*}_j \mathbf{5}_j$. In the flipped 
$SU(5)\times U(1)_X$ models, we also have two kinds of operators
$\mathbf{(10, 1)^*}_i \mathbf{(10, 1)}_i \mathbf{(10, 1)^*}_j \mathbf{(10, 1)}_j$
and $\mathbf{(10, 1)^*}_i \mathbf{(10, 1)}_i \mathbf{(\overline{5}, -3)^*}_j 
\mathbf{(\overline{5}, -3)}_j$.
In $SO(10)$ models, we only have one kind of operators 
 $\mathbf{16^*}_i \mathbf{16}_i \mathbf{16^*}_j \mathbf{16}_j$.
In terms of the SM fields, we obtain
the possible dimension-six operators which contribute to the 
proton decay~\cite{Nath:2006ut}
\begin{eqnarray}
\label{O1}
\textit{O}_I&=& {{g_U^2}\over {2M^2_{(X,Y)}}} \ \overline{U_i^c} \ \gamma^{\mu}  \ Q_i 
\ \overline{E_j^c} \gamma_{\mu} Q_j~,~\,
\\ 
\label{O2}
\textit{O}_{II}&=& {{g_U^2}\over {2M^2_{(X,Y)}}}
\ \overline{U_{i}^c} \ \gamma^{\mu} \ {Q_{i}}   \
\overline{D^c_{j}} \ \gamma_{\mu} \ {L_{j }} ~,~\, \\
\label{O3}
\textit{O}_{III}&=& {{g_U^2}\over {2M^2_{(X^{'},Y^{'})}}}
\ \overline{D_{i }^c} \ \gamma^{\mu} \ {Q_{i}}  \
\overline{U_{j}^c} \ \gamma_{\mu} \ {L_{j}} ~,~\, \\
\label{O4}
\textit{O}_{IV}&=&  {{g_U^2}\over {2M^2_{(X^{'},Y^{'})}}}
\ \overline{D^C_{i}} \ \gamma^{\mu} \ {Q_{i }}    \
\overline{N_j^c}_L \ \gamma_{\mu} \ {Q_{j}} ~,~\,
\end{eqnarray}
where  $g_U$ is the unified gauge coupling at the GUT scale,
and $M_{(X,Y)}$ and $M_{(X^{'},Y^{'})}$ 
 are the masses of the superheavy gauge bosons in the 
$SU(5)$ models and flipped $SU(5)\times U(1)_X$ models, respectively. 
In the $SU(5)$ models, we obtain the 
 effective operators $\textit{O}_I$ and
$\textit{O}_{II}$ respectively in Eqs.~(\ref{O1}) and~(\ref{O2}) 
 after the superheavy gauge fields 
$(X, Y)=\mathbf{({3},{ 2},5/6)}$ are integrated out.
In the flipped $SU(5)\times U(1)_X$ models, we obtain the effective
operators $\textit{O}_{III}$ and $\textit{O}_{IV}$ 
respectively in Eqs.~(\ref{O3}) and~(\ref{O4}) after 
 the superheavy gauge fields $(X^{'}, Y^{'})=\mathbf{({3},{2},-1/6)}$ 
 are integrated out.
Because both the $SU(5)$ models and the flipped $SU(5)\times U(1)_X$ models
can be embedded into the $SO(10)$ models, we have all these superheavy
gauge fields as well as all the above dimension-six proton
decay operators. Note that the dimension-six 
proton decays  have not been observed from the experiments, 
we obtain that the GUT 
scale is higher than about $5\times 10^{15}$ GeV.
Because the GUT scale in our models can be as small as 
$5.3\times10^{13}~{\rm GeV}$, we require that
the $(X,Y)$ gauge bosons in the $SU(5)$ models and
the $(X,Y)$ and $(X^{'}, Y^{'})$ gauge bosons in the
$SO(10)$ models do not generate  the above dimension-six proton decay
operators. Therefore, we need to forbid at least some of the couplings
between the superheavy gauge
fields and the SM fermions in the model building.

Second, let us consider the F-theory GUTs 
which do not have proton decay problem.
In the F-theory $SU(5)$ model proposed in Ref.~\cite{Li:2009cy}, the
Higgs fields $\mathbf{{5}}_H=(T_u,~H_u)$ and 
$\mathbf{\overline{5}}_{\overline{H}}=(T_d,~H_d)$ are on the 
different Higgs curves, and $T_u$ and $T_d$ do not have zero modes
by choosing proper $U(1)$ fluxes. 
And then the KK modes of $T_u$ and $T_d$ do not 
form vector-like particles, {\it i.e.}, the third term
in Eq.~(\ref{PD-D5}) does not exist.  The mass terms between the KK modes of
$T_u$ and $T_d$ arise from the usual $\mu$ term. So the proton partial lifetime via
the dimension-five proton
decay is proportional to $M^2_{T_u} M^2_{T_d} M^2_S/\mu^2$.
In generic GUTs with high-scale supersymmetry breaking, we have $M_S \simeq \mu$,
and $M_{T_u} \sim M_{T_d} \sim M_U$.
Thus, the  proton partial lifetime via
the dimension-five proton
decay is proportional to $M_U^2 \ge 10^{26}~{\rm GeV}^2$, which is much
larger than $10^{20}~{\rm GeV}^2$. And then we do not
have the dimension-five proton decay problem. Moreover, the SM quarks
$Q_i$ and $U_i^c$ are on different matter curves. And then the
$X$ and $Y$ gauge bosons can not couple to both $Q_i$ and $U_i^c$.
Therefore, we do not have the dimension-six proton decay problem
via superheavy gauge boson exchanges.

In the Type I and Type II F-theory $SO(10)$ models proposed 
in Ref.~\cite{Li:2009cy}
where the $SO(10)$ gauge symmetry is broken down to the 
$SU(3)_C\times SU(2)_L \times SU(2)_R \times U(1)_{B-L}$ gauge symmetry, 
the SM fermions $Q_i$, $E_i^c$, and $N_i^c$ are on one matter curve,
while $U_i^c$, $D_i^c$, and $L_i$ are on the other matter curve.
On the Higgs $\mathbf{10}_H=(T_u,~H_u,~T_d,~H_d)$ curve, 
$T_u$ and $T_d$ do not have zero modes
by choosing proper $U(1)$ fluxes, and the KK modes of $T_u$ and $T_d$ do not 
form vector-like particles. Thus, similar to the discussions in
the above F-theory $SU(5)$ models, we do not
have the dimension-five proton decay problem. Moreover, the SM quarks
$Q_i$ and $U_i^c$/$D_i^c$ are on different matter curves. So the
$X$ and $Y$ gauge bosons can not couple to both $Q_i$ and $U_i^c$,
and the  $X'$ and $Y'$ gauge bosons can not couple to both $Q_i$ and $D_i^c$,
Therefore, we do not have the dimension-six proton decay problem
via superheavy gauge boson exchanges.

Third, we consider the five-dimensional orbifold $SU(5)$ and
$SO(10)$ models on $S^1/(Z_2\times Z_2')$ where the proton decay problems can
be solved 
as well~\cite{kawa, GAFF, LHYN, AHJMR, Li:2001qs, Dermisek:2001hp, Li:2001tx}.
We assume that the fifth dimension is a circle $S^1$ with  coordinate $y$
and radius $R$.
The orbifold $S^1/(Z_2 \times Z_2')$ is obtained by the circle $S^1$ moduloing 
the following equivalent classes
\begin{eqnarray}
P:~~~  y \sim -y~,~~~~~~~~~~P':~~~y' \sim -y'~,~\,
\end{eqnarray}
where $y'=y+\pi R/ 2$. There are two inequivalent 3-branes located at
the fixed points $y=0$ and $y=\pi R/2$, which are denoted by $O_B$ and $O'_B$,
 respectively. 
In particular, the zero modes of the SM fermions in the bulk
do not form the complete GUT representations due to the orbifold
gauge symmetry breaking~\cite{Li:2001tx}.

In the orbifold $SU(5)$ models (for a concrete example,
 see Ref.~\cite{LHYN}), 
the $SU(5)$ gauge symmetry
is broken down to the SM gauge symmetry via orbifold projections.
With suitable representations for the $Z_2$ and $Z'_2$ parities,
the $SU(5)$ gauge symmetry is preserved on the 
$O_B$ 3-brane, while it is broken down to the SM gauge symmetry
on the $O'_B$ 3-brane. To realize the gauge coupling relation 
in Eq.~(\ref{GCRelation}), we introduce the adjoint Higgs field in the 
$\mathbf{24}$ representation on the  $O_B$ 3-brane. 
The gauge coupling relation 
in Eq.~(\ref{GCRelation}) can be generated via 
the suitable dimension-five
operators after the adjoint Higgs field acquires 
the VEV~\cite{Hill:1983xh, Shafi:1983gz, Ellis:1985jn, Li:2010xr}.
We put the Higgs fields $\mathbf{{5}}_H=(T_u,~H_u)$ and 
$\mathbf{\overline{5}}_{\overline{H}}=(T_d,~H_d)$ in the bulk,
and then $T_u$ and $T_d$ do not have zero modes due to the
orbifold projections. In particular, the KK modes
for $T_u$ and $T_d$ only have vector-like mass term via $\mu$ term.
Thus, similar to  the discussions in the above F-theory GUTs, 
we do not have the dimension-five proton
decay problem. To forbid the dimension-six proton decay, 
we put the SM fermion superfields $\mathbf{10}_i$ 
and $\mathbf{10}'_i$ in the bulk
with suitable $Z_2$ and $Z_2'$
parity assignments where $i=1,~2,~3$. 
We obtain the SM fermions $Q_i$ as zero modes from $\mathbf{10}_i$ while
we obtain the SM fermions $U_i^c$ and $E_i^c$  as zero modes 
from $\mathbf{10}'_i$. Because the
$X$ and $Y$ gauge bosons can not couple to both $Q_i$ and $U_i^c$,
 we do not have the dimension-six proton decay problem
via superheavy gauge boson exchanges.

In the orbifold $SO(10)$ models (for a concrete example,
 see Ref.~\cite{Dermisek:2001hp}), 
the $SO(10)$ gauge symmetry is broken down to the Pati-Salam
$SU(4)_C\times SU(2)_L\times SU(2)_R$ 
gauge symmetry via orbifold projections.
With suitable representations for the $Z_2$ and $Z'_2$ parities,
the $SO(10)$ gauge symmetry is preserved on the 
$O_B$ 3-brane, while it is broken down to the Pati-Salam gauge symmetry
on the $O'_B$ 3-brane. To realize the gauge coupling relation 
in Eq.~(\ref{GCRelation}), we introduce the symmetric Higgs field in the 
$\mathbf{54}$ representation on the  $O_B$ 3-brane. 
The gauge coupling relation 
in Eq.~(\ref{GCRelation}) can be generated via the suitable dimension-five
operators after the symmetric Higgs field acquires 
the VEV~\cite{Hill:1983xh, Shafi:1983gz, Ellis:1985jn, Li:2010xr}.
We put the Higgs field $\mathbf{10}_H=(T_u,~H_u,~T_d,~H_d)$ in the bulk,
and then $T_u$ and $T_d$ do not have zero modes due to
orbifold projections. In particular, the KK modes
for $T_u$ and $T_d$ only have vector-like mass term via $\mu$ term.
Thus, similar to  the discussions in the above F-theory GUTs
 and the orbifold $SU(5)$ models, we do not have 
the dimension-five proton
decay problem. To forbid the dimension-six proton decay, 
we put the SM fermion superfields $\mathbf{16}_i$ 
and $\mathbf{16}'_i$ in the bulk
with suitable $Z_2$ and $Z_2'$ parity assignments where $i=1,~2,~3$. 
We obtain the left-handed SM fermions $Q_i$ and $L_i$ 
as zero modes from $\mathbf{16}_i$ while 
we obtain the right-handed SM fermions
$U_i^c$, $D_i^c$, $N_i^c$ and $E_i^c$ as zero modes 
from $\mathbf{16}'_i$.  Because the
$X$ and $Y$ gauge bosons can not couple to both $Q_i$ and $U_i^c$
and the $X'$ and $Y'$ gauge bosons can not couple to  
both $Q_i$ and $D_i^c$,
 we do not have the dimension-six proton decay problem
via superheavy gauge boson exchanges.

Fourth, let us comment on the superheavy threshold corrections 
on the gauge coupling unification in our models.
In the F-theory $SU(5)$ and $SO(10)$ models, we shall
have the superheavy threshold corrections from
the Kaluza-Klein (KK) modes and heavy string
modes. Because our unification scale is smaller
than or equal to $10^{16}~{\rm GeV}$, we do not have string
theshfold corrections since the string scale is generic around 
$4\times 10^{17}~{\rm GeV}$. Also, the
KK modes can have masses around the GUT scale or higher, and
then their effects on the gauge coupling unification can be negligible
as well. Moreover, in the orbifold $SU(5)$ and $SO(10)$ models,
 we shall  have the superheavy threshold corrections from the
 KK modes. Because the masses of the KK modes can not be larger
than the GUT scale, we might have appreciable threshold corrections
on the gauge coupling unification, which definitely
deserves further detailed study. Thus, for simplicity, we assume that
the KK mass scale is equal to the GUT scale in this paper.



\section{Conclusion}


Inspired by the string landscape and 
the unified gauge coupling relation in 
 the F-theory GUTs and GUTs with suitable high-dimensional
operators, we studied the 
canonical gauge coupling unification in the SM with
high-scale supersymmetry breaking. In the SM with
GUT-scale supersymmetry breaking,  the
 gauge coupling unification can be achieved
 at about $5.3\times 10^{13}~{\rm GeV}$, and the
Higgs boson mass is predicted to range from 130 GeV to 147 GeV.
In the SM with supersymmetry breaking scale from $10^4~{\rm GeV}$
to $5.3\times 10^{13}~{\rm GeV}$,  gauge coupling unification
can always be realized, and the corresponding
GUT scale $M_U$ is  from $10^{16}~{\rm GeV}$ to 
$5.3\times 10^{13}~{\rm GeV}$, respectively. Also,
we obtained the Higgs
boson mass from 114.4 GeV to $147$ GeV.
Moreover, the discrepancies among the SM gauge couplings
at the GUT scale are less than about 4-6\%.
Furthermore, we presented the $SU(5)$ and $SO(10)$ models from the 
F-theory model building and orbifold constructions, and showed
that there are no dimension-five and dimension-six
proton decay problems even if 
$M_U \le 5\times 10^{15}~{\rm GeV}$.

\begin{acknowledgments}


This research was supported in part 
by the Natural Science Foundation of China 
under grant numbers 10821504 and 11075194 (TL),
and by the DOE grant DE-FG03-95-Er-40917 (TL and DVN).

\end{acknowledgments}


\end{document}